\documentclass[review]{elsarticle}
\makeatletter
\def\ps@pprintTitle{%
 \let\@oddhead\@empty
 \let\@evenhead\@empty
 \def\@oddfoot{}%
 \let\@evenfoot\@oddfoot}
\makeatother
\usepackage{hyperref,graphicx}
\usepackage[normalem]{ulem}
\usepackage[margin=1.0in]{geometry} 
\usepackage{multirow}
\usepackage{subfigure}
\usepackage{amsfonts}
\usepackage{amsmath,bm}
\usepackage{booktabs, threeparttable}
\usepackage{array}
\newcolumntype{L}[1]{>{\raggedright\arraybackslash}p{#1}}
\DeclareMathOperator{\Tr}{Tr}
\usepackage{color}

\begin{document}

\title{Physical mechanisms of the Soret effect in binary Lennard-Jones liquids elucidated with thermal-response calculations}

\author[ucfaddress,AMPACaddress,REACT]{Patrick K. Schelling\corref{mycorrespondingauthor}}
\cortext[mycorrespondingauthor]{Corresponding author.}
\ead{patrick.schelling@ucf.edu}
\address[ucfaddress]{Department of Physics, University of Central Florida, Orlando, FL 32816-2385, USA}
\address[AMPACaddress]{Advanced Materials Processing and Analysis Center, University of Central Florida, Orlando, FL 32816-2385, USA}
\address[REACT] {Renewable Energy and Chemical Transformations (REACT) Cluster, University of Central Florida, Orlando, FL 32816-2385, USA}

\begin{abstract}
The Soret effect is the tendency of fluid mixtures to exhibit concentration gradients in the presence of a temperature gradient. 
Using molecular-dynamics simulation of two-component Lennard-Jones liquids, it is demonstrated that spatially-sinusoidal heat pulses generate both temperature and pressure gradients. Over short timescales, the dominant effect is the generation of compressional waves which dissipate over time as the system approaches mechanical equilibrium. The approach to mechanical equilibrium is also characterized by a decrease in particle density in the high-temperature region and an increase in particle density in the low-temperature region. It is demonstrated that concentration gradients develop rapidly during the propagation of compressional waves through the liquid. Over longer timescales, heat conduction occurs to return the system to thermal equilibrium, with the particle current acting to restore a more uniform particle density.
It is shown that the Soret effect arises due to the fact that the two components of the fluid exhibit a different response to pressure gradients. First, the so-called isotope effect occurs because light atoms tend to respond more rapidly to the evolving conditions.  In this case, there appears to be a connection to previous observations of ``fast sound'' in binary fluids. Second, it is shown that the partial pressures of the two components in equilibrium, and more directly the relative magnitudes of their derivatives with respect to temperature and density, determine which species accumulates in the high- and low-temperature regions. In the conditions simulated here, the dependence of the partial pressure on density gradients is larger than the dependence on temperature gradients. This is directly connected to accumulation of the species with the largest partial pressure in the high-temperature region, and an accumulation of the species with the smallest partial pressure in the low-temperature region. The results suggest that further development of theoretical descriptions of the Soret effect might begin with hydrodynamical equations in  two-component liquids. 
Finally, it is suggested that the recently proposed concept of ``thermophobicity'' may be related to the sensitivity of partial pressures in a multicomponent fluid to changes in temperature and density.

 \end{abstract}


\maketitle

\section{Introduction}
The Soret effect describes the tendency of concentration gradients to develop in the presence of external heat sources. In a binary fluid, this can be observed as a tendency of one component to accumulate in high-temperature region, while the other component accumulates in the low-temperature region. The phenomenological equation which relates to the particle flux density $\bm{J}_{1}$ of component 1 to the composition and temperature gradients is\cite{ludwig_1856,soret_1879,Morozov:2009wa},
\begin{equation}
\bm{J}_{1} = -D_{12} \rho \left [\bm{\nabla} w_{1} + S_{T} w_{1}(1 - w_{1}) \bm{\nabla} T \right] ,
\end{equation}
in which $w_{1}$ is the weight fraction of component $1$ and $T$ is the local temperature. The interdiffusion coefficient $D_{12}$ and the Soret coefficient $S_{T}$ determine the response of the system to gradients in composition and temperature respectively. In the presence of a steady-state heat current, the flux density $\bm{J}_{1}=0$ and the Soret coefficient  can be determined by either measurement or computation of both the stationary composition and temperature gradients\cite{Reith_2000}.

While the Soret effect has been studied for many decades, a simple and coherent physical description of the effect in multicomponent liquids has not yet emerged. However, several general trends have been established. It has been suggested that the effect can be separated into an ``isotopic'' and ``chemical'' effect which are additive\cite{Hoang_2022}. The isotopic effect arises entirely due to mass differences, whereas the chemical effect depends on many factors including molecule size and interaction strength. Moreover, an extensive study of equimolar mixtures of organic liquids has revealed that a pure component property termed ``thermophobicity'' is predictive of the behavior of the Soret coefficient of mixtures\cite{Hartmann:2012va,Hartmann_2014}. Theoretical interpretation of these results was based on the ``mechanical'' model developed by Morozov\cite{Morozov:2009wa}.

Molecular-dynamics (MD) simulations have been used to elucidate the Soret effect. There are generally two different MD approaches which have been used. One approach is to mimic experiment by generating an explicit temperature gradient\cite{Reith_2000}. Once steady-state has been achieved, it is very straightforward to compute an average concentration gradient and temperature gradient. For example, this approach was used in a rather extensive set of simulations of two-component Lennard-Jones (LJ) liquids\cite{Reith_2000}, and many other similar simulations using LJ potentials have also been reported\cite{SIMON1998151,Perronace:2002aa,Galli_ro_2003,YEGANEGI2005171}. Another related approach is to drive currents using an external field, and then determine the Onsager transport coefficients via calculation of the resulting currents\cite{Evans_1991,Perronace:2002aa}. The second widely-used approach is equilibrium MD simulation analyzed using Green-Kubo (GK) theory to compute the Onsager transport coefficients\cite{Heyes_1992}. For example, there are many studies of binary liquid metal alloys using this approach\cite{Evteev_2014,TUCKER201654}. Recently, important insight into the Soret effect was obtained in a recent MD study by Hafskjold using an innovative approach of computing the transient response to a heat perturbation\cite{Hafskjold_2017}. Specifically, by applying a heat pulse to a binary Lennard-Jones mixture, it was demonstrated that the primary result is a shock/pressure which propagates through the system. This result was to be expected. However, the interesting and perhaps surprising observation was made that the interdiffusion occurs rapidly with concentration gradients developing over timescales comparable to the period of the compression waves\cite{Hafskjold_2017}. Additional work by Bonella and coworkers\cite{Bonella_2017,Ferrario_2016} also made similar observations related to the approach to steady state. These previous works indicate an important connection between density and pressure fluctuations, sound waves, and the Soret effect which is yet to be completely elucidated. 

In this paper additional insight into the observations first reported in  \cite{Hafskjold_2017} and \cite{Bonella_2017,Ferrario_2016} is established. The general approach of determining the response to a perturbation is formalized based on thermal response functions first applied to heat transport in LJ solids\cite{Fernando_2020}, and more recently used to analyze anomalous heat transport in one-dimensional chains\cite{Bohm:2022aa}. In this approach, an input heat pulse generates a strong pressure gradient with a resulting compressional wave. During this phase, particles of both species are driven away from the hot region as expected. This is followed by a slower process which involves transport of heat. As heat transport occurs and thermal equilibrium is gradually attained, mass tends to flow back into the depleted regions. In fact, both components flow opposite to the direction of the heat current. This latter phase appears to be primarily where the separation occurs. The details of the process are determined by two factors. First, atoms with smaller masses respond more rapidly to changing conditions, and hence accumulate in the hot region. This is consistent with previous observations of the isotope effect. Second, the importance of the partial pressures of the two components is revealed. Specifically, the driving force for the species with the most positive partial pressure tends to be larger, and hence tends to accumulate in the hot region. Consistent with the observations by in \cite{Hafskjold_2017}, the separation tends to occur on a timescale consistent with the period of compressional waves, indicating a ``mechanical'' mechanism for separation.

\section{Theory and Methodology}

In a previous work\cite{Reith_2000}, non-equilibrium MD was used to determine concentration gradients in two-component LJ mixtures that result from an imposed temperature gradient. The pairwise interaction potential between particles of species $s$ and $s^{\prime}$ is given in the LJ model by,
\begin{equation}
u_{ss^{\prime}}^{0}(r)=-4\epsilon_{ss^{\prime}} \left[\left(\sigma_{ss^{\prime}} \over r \right)^{6} -\left(\sigma_{ss^{\prime}} \over r \right)^{12} \right] .
\end{equation}
Several different sets of LJ parameters were used to establish systematic behavior. Component $s=2$ was always described by parameters consistent with Ar. This choice was made to compare directly to previous results for the Soret effect\cite{Reith_2000}. Specifically,  $\epsilon_{22} = k_{B}T_{Ar}$ with $T_{Ar}=120K$ and $\sigma_{22}=3.4\AA$ were chosen. The mass of component $s=2$ was taken to be $m_{Ar}=39.948$ in atomic mass units. Following the previous study\cite{Reith_2000}, different parameter sets were used for component $s=1$. Specifically, the effects of varying the mass ratio ${m_{1} \over m_{2}}$, interaction strength ${\epsilon_{11} \over \epsilon_{22}}$, and particle size ${\sigma_{11} \over \sigma_{22}}$ will be assessed independently.
For the two-component systems, interactions between the two species were specified by the Lorenz-Berthelot mixing rule,
\begin{equation}
\epsilon_{12}=\epsilon_{21} = \sqrt{\epsilon_{11}\epsilon_{22}}
\end{equation}
\begin{equation}
\sigma_{12}=\sigma_{21} = {1 \over 2} \left( \sigma_{11}+\sigma_{22} \right) .
\end{equation}
To ensure computational efficiency, a cutoff of $r_{c}=3 \sigma_{11}$ was used, with $\sigma_{11}$ always taken to be the size of the largest atom. For atom separations $r>r_{c}$, the interaction was taken to be zero. When $r<r_{c}$, to ensure no discontinuities in the potential or forces existed, the potentials were smoothed by calculating the interactions from the effective potential,
\begin{equation}
u_{ss^{\prime}}(r)=u_{ss^{\prime}}^{0}(r)-u_{ss^{\prime}}^{0}(r=r_{c})-(r-r_{c})\left[du_{ss^{\prime}} \over dr\right]_{r=r_{c}} .
\end{equation}
Because the cutoff $r_{c}=3\sigma_{11}$ is quite large, the effect of smoothing is minimal for separations near the equilibrium bond length. After smoothing, the leading discontinuities correspond to second-order derivatives of the potential energy function.

To compute partial pressures, the standard Parrinello-Rahman approach was used\cite{Parrinello_1981}. Specifically, the  stress tensor is computed using,
\begin{equation}
\boldsymbol{\pi} \Omega  = \sum_{s=1}^{2}\sum_{i=1}^{N_{s}} m_{s} \langle \bm{v}_{si} \bm{v}_{si} \rangle + {1 \over 2}\sum_{s=1}^{2}\sum_{i=1}^{N_{s}}  \sum_{s^{\prime}j\ne s i} \langle \bm{F}_{si,s^{\prime} j} \bm{r}_{si,s^{\prime} j}  \rangle .
\end{equation}
The pressure is determined by the trace of the quantity above, namely $p={1 \over 3} \Tr{\left[ \boldsymbol{\pi} \right]}$. To define partial pressures, summation occurs only over one species. The partial stress tensor $\bm{\pi}^{(s)}$ is defined accordingly,
\begin{equation}
\bm{\pi}^{(s)}\Omega  = \sum_{i=1}^{N_{s}} m_{s} \langle \bm{v}_{si} \bm{v}_{si} \rangle + {1 \over 2}\sum_{i=1}^{N_{s}}\sum_{s^{\prime}j\ne s i} \langle \bm{F}_{si,s^{\prime}j} \bm{r}_{si,s^{\prime}j}  \rangle .
\end{equation} 
Then the partial stresses are given by the trace $p_{s}= {1 \over 3} \Tr{\left[ \boldsymbol{\pi}^{(s)} \right]}$.
For a two component system $s=1,2$ then, the relationship between the total pressure and the partial pressures is
\begin{equation}
p = p_{1} + p_{2} .
\end{equation}
This may not be a precise thermodynamic description of partial pressure, which should be related to the fugacities and hence the chemical potentials of the different components. However, this definition is consistent with the definition used in the Bearman-Kirkwood theory for the statistical mechanics of transport processes in multicomponent systems\cite{Bearman_1958}. Hence, this definition is particularly relevant for understanding hydrodynamics of a multicomponent fluid\cite{Bearman_1958}. Effectively the same form was also used by Morozov in his more recent work on thermodiffusion\cite{Morozov:2009wa}.

Here the partial pressures of the two species are determined for three different LJ binary liquids. In addition, derivatives of $p_{s}$ with respect to temperature and density were also computed. Calculations were performed for a cubic supercell with side length $L$ and volume $\Omega=L^{3}$ and a total of $N=4000$ particles. Some calculations with larger cell sizes were performed as noted later. Periodic boundary conditions were applied in each direction. The systems considered correspond to equimolar mixtures with $N_{1}=N_{2}=2000$. The simulation volume was held constant during integration. Integration was performed using the velocity Verlet algorithm with a timestep $dt=10^{-3}$ in reduced units. 
The quantities $p_{s}$ and their temperature and concentration partial derivatives were determined by varying both concentration and temperature. Specifically, for each simulation system, the reduced temperature was varied by about $\pm 2\%$ to obtain $\left({\partial p_{s} \over \partial T}\right)_{n_{1},n_{2}}$. To obtain concentration dependence, the number of atoms of one species was varied by $\pm 20$. These calculations were used to determined the quantities $\left({\partial p_{s} \over \partial T}\right)_{T,n_{s^{\prime}}}$. These results are accumulated in Table \ref{table1}. Also shown in Table \ref{table1} are the self-diffusion coefficients $D_{1}$ and $D_{2}$ for the two species obtained from the mean-squared displacement. With the exception of the results for ${\sigma_{11} \over \sigma_{22}}=1.9$, the values for $D_{1}$ and $D_{2}$  are in good agreement with those reported previously\cite{Reith_2000}.  For the  ${\sigma_{11} \over \sigma_{22}}=1.9$ results, the current simulations predict consistently smaller values than those in Ref.\cite{Reith_2000}. Finally, the simulated reduced temperature was $T^{*} = {k_{B}T \over \epsilon_{12}}=0.85$ and the reduced particle density was $n^{*}={N \sigma_{12}^{3} \over \Omega}=0.81$. This state point was chosen to agree with the previously-reported simulations by Reith and M\"{u}ller-Plathe \cite{Reith_2000}. This state point was also the reference point chosen for the partial derivatives in Table \ref{table1}.

\begin{table}
\begin{center}
\caption{Tabulated results for quantities relevant to generalized thermodynamic forces. Energies $p_{1}\Omega$ and $p_{2}\Omega$ are given in units $\epsilon_{11}$. Temperature derivatives are taken with respect to the reduced temperature defined by $T^{*}={k_{B}T \over \epsilon_{11}}$. Previously reported results for $D_{1}$ and $D_{2}$ from \cite{Reith_2000} are shown in parentheses. Partial derivatives of $p_{1}$ and $p_{2}$ assumed both were only functions of $T$, $n_{1}$, and $n_{2}$.
}
\begin{tabular} {|c|c|c|c|c|c|c|c|c|c|c|}
\hline
System & $ {p_{1} \over n_{1}}$ & ${p_{2} \over n_{2}}$ & $ {1 \over n_{1}}{\partial p_{1} \over \partial T^{*}}$ &   ${1 \over n_{2}}{\partial p_{2}  \over \partial T^{*}}$  & ${\partial p_{1} \over \partial n_{1}}$   & ${\partial p_{2} \over \partial n_{1}}$ &  ${\partial p_{1} \over \partial n_{2}}$ &   ${\partial p_{2} \over \partial n_{2}}$ & $D_{1}\times 10^{5}$ cm$^{2}$s$^{-1}$ &
$D_{2}\times 10^{5}$ cm$^{2}$s$^{-1}$\\\hline 
${m_{1} \over m_{2}} = 8$ & 1.36 & 1.36 & 5.23  & 5.23 & 7.21  & 5.72  & 5.72 & 7.21  & 1.51 (1.51)& 1.70 (1.79)\\\hline
${\epsilon_{1} \over \epsilon_{2}} = 2.5$ &  0.14 & 1.27 & 4.85    & 5.26  &  3.57   & 3.53& 5.42  & 7.33 & 3.13 (3.05) & 3.90 (4.00)\\\hline
${\sigma_{1} \over \sigma_{2}} = 1.9$& 10.69 & 2.92 & 11.59    &  3.93 & 73.57  & 16.45 & 11.96 & 6.42 & 0.38 (0.58) & 1.22 (1.84) \\\hline
\end{tabular}
\label{table1}
\end{center}
\end{table}

The microscopic definitions of the local particle current densities for the two species are given by,
\begin{equation}
\bm{J}_{s}(\bm{r},t) = \sum_{j=1}^{N_{s}} \bm{v}_{sj}(t) \delta^{(3)} (\bm{r}-\bm{r}_{sj}).
\end{equation}
The particle densities for the two species are defined by,
\begin{equation} \label{dens}
n_{s}(\bm{r}) = \sum_{j=1}^{N_{s}} \delta^{(3)} (\bm{r}-\bm{r}_{sj}).
\end{equation}
in which $\vec{v}_{sj}$ represents the velocity vector and $\vec{r}_{sj}$ the position vector of particle $j$ within species $s$, and $N_{s}$ represents the total number of particles of species $s$. 
The Fourier transforms of the particle current densities are represented by,
\begin{equation} \label{js}
\tilde{J}_{s}(\bm{k},t) ={1 \over \Omega}  \sum_{j=1}^{N_{s}} { \bm{v}_{sj}(t) \cdot \bm{k} \over |\bm{k}|} e^{-i \bm{k} \cdot \bm{r}_{sj}(t)} ,
\end{equation}
in which use that longitudinal currents are relevant for the development of density and concentration gradients.

Next, the usual assumption of linear-response is applied so that the current densities which result from an external heat input $ u^{(ext)}(\bm{r}^{\prime})$ at $t=0$ are described by the expressions,
\begin{equation} \label{jqresp}
\bm{J}_{s}(\bm{r}, \tau) =  -{1 \over \Omega} \int_{\Omega} K_{sQ}(\bm{r}-\bm{r}^{\prime},\tau) \bm{\nabla} u^{(ext)}(\bm{r}^{\prime}) d^{3}r^{\prime} .
\end{equation}
In other words, heat input from an external source at $t=0$ into the two-component liquid results in current densities at later times $\tau>0$. 
This is analogous to our previous work in Ref. \cite{Fernando_2020}, but here applied to particle current densities. 
Also, as noted earlier, Hafskjold has also reported responses to external heat inputs in two-component LJ liquids, although he did not consider a specific definition of a response function\cite{Hafskjold_2017}. It might also be noted that the idea of computing transport coefficients using perturbations due to an external field has already been used in the context of transport in liquids\cite{Perronace:2002aa,Evans_1991}. The relationship between previously used approaches and the one considered here is not immediately obvious, but one distinction is that the approach considered here involves spatially-varying perturbations that permit calculations in reciprocal space.

As with our previous work\cite{Fernando_2020}, it is useful in periodic structures to consider the linear response equations in reciprocal space. Here, while there is no crystal lattice, the system is still subject to periodic-boundary conditions. Hence, we expand the external heat perturbation as,
\begin{equation}
 u^{(ext)}(\bm{r}^{\prime}) = \sum_{\bm{k}} \tilde{u}^{(ext)}(\bm{k}) e^{i \bm{k} \cdot \bm{r}^{\prime}} 
\end{equation}
\begin{equation}
\bm{\nabla} u^{(ext)}(\bm{r}^{\prime}) =
i\bm{k} \sum_{\bm{k}} \tilde{u}^{(ext)}(\bm{k}) e^{i \bm{k} \cdot \bm{r}^{\prime}} ,
\end{equation}
in which $\bm{k}$ represent the possible reciprocal lattice vectors of the simulation supercell.
Since the system is isotropic, we can omit vector notation and assume that current responses are always parallel to the wave vector $\bm{k}$ of the perturbation, and then,
\begin{equation}
\tilde{J}_{s} (k,\tau)=  -ik \tilde{K}_{sQ}(k,\tau)\tilde{u}^{(ext)}(k) .
\end{equation}
The approach we take is the impose an energy perturbation at $t=0$ with reciprocal space component $\tilde{u}^{(ext)}(k)$, and then compute an ensemble averaged current responses $\tilde{J}_{1} (k,\tau)$ and $\tilde{J}_{2} (k,\tau)$
Using the equations above, the three response functions can be determined,
\begin{equation} \label{respcalc}
\tilde{K}_{sQ}(k,\tau) = {i\langle \tilde{J}_{s} (k,\tau) \tilde{u}^{(ext)}(-k) \rangle \over k \langle \tilde{u}^{(ext)}(k)\tilde{u}^{(ext)}(-k)\rangle}  .
\end{equation}
In this equation, the angle brackets are meant to represent an nonequilibrium ensemble average over many realizations of the external perturbation. This quantity is to be computed for both components of the liquid $s=1,2$.

The liquid was generated by melting an initial randomized structure with atoms of both sites situated at FCC lattice sites.
After equilibration, the external perturbation is implemented by scaling instantaneous velocities of each particle $j$ of species $s$ according to,
\begin{equation}\label{pert}
\bm{v}_{sj}(0) \rightarrow \bm{v}_{sj}(0) \sqrt{1 + b \cos{\left(k z_{sj}\right)}} ,
\end{equation}
in which the perturbation is applied at time $t=0$ along the $\hat{z}$ direction and $\vec{k}\cdot \vec{r}_{sj}=kz_{sj}$. Unless otherwise noted,
the amplitude of the perturbation was taken to be $b=0.30$.  Given the perturbation imposed as described by Eq. \ref{pert}, it can be easily demonstrated that
in real space, the external perturbation is simply,
\begin{equation}\label{pertrs}
u^{(ext)} (z) = {3 Nb k_{B}T \over 2\Omega} \cos{\left(kz \right)} .
\end{equation} 
Consequently, the terms related to the external source in Eq. \ref{respcalc} are given by,
\begin{equation} \label{pertrecip}
 \tilde{u}^{(ext)}(k) = \tilde{u}^{(ext)}(-k)= {3Nb k_{B}T \over 4 \Omega} ,
\end{equation}
in which $k = {2 \pi \over L}$.
Given the particular perturbation in Eq. \ref{pertrs}, the response function is computed using Eq. \ref{respcalc}. Finally, the real-space heat flux for this perturbation is given by,
\begin{equation}
J_{s}(z,t) = \left({3Nb k_{B}T \over 2\Omega} \right) k \tilde{K}_{sQ}(k,t)\sin{\left(kz\right)} .
\end{equation}
With this heat flux, after application of the continuity equation and integration over time, 
 the real-space number densities $n_{1}(z,\tau)$ and  $n_{2}(z,\tau)$ for the two components,
\begin{equation} \label{dist1}
n_{s}(z,\tau)-{N_{s} \over \Omega}=-\left({3Nb k_{B}T \over 2 \Omega} \right) k^{2} \cos{\left(kz\right)}\int_{0}^{\tau}\tilde{K}_{sQ}(k,t)dt=A_{s}(k,\tau) \cos(kz),
\end{equation}
are obtained.
Comparison to the expression for the external heat pulse, these can be also written,
\begin{equation}
n_{s}(z,\tau)= \left[-k^{2} \int_{0}^{\tau}\tilde{K}_{sQ}(k,t)dt \right] u^{(ext)}(z) .
\end{equation}

\section{Results}

\begin{figure}
\begin{centering}
\includegraphics[width=0.5\textwidth]{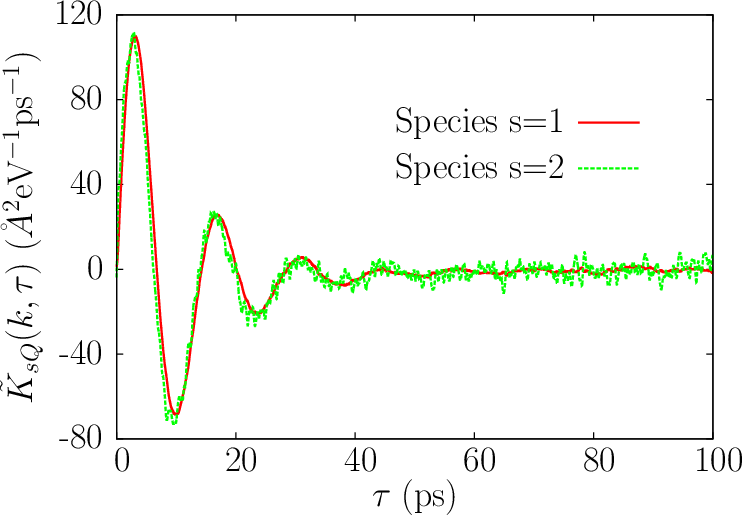} 
\caption{Response functions $\tilde{K}_{sQ}(k,\tau)$ obtained for an external perturbation at time zero for the system plotted for both species $s=1$ and $s=2$ with mass ratio ${m_{1} \over m_{2}}=8$. The perturbation and response correspond to $k={2 \pi \over L}$.  The parameter $b=0.30$ for the perturbation strength was used.
}
\label{ksq}
\end{centering}
\end{figure}

First we consider the isotopic effect for a system with mass ratio ${m_{1} \over m_{2}}=8$. The other parameters  corresponded to ${\epsilon_{11}\over \epsilon_{22}}=1$ and ${\sigma_{11} \over \sigma_{22}}=1$. The supercell size, number of atoms $N_{1}=N_{2}=2000$, and other conditions were the same as those used to compute the values in Table \ref{table1}.
In Fig. \ref{ksq} the ensemble-averaged response functions $\tilde{K}_{sQ}(k,\tau)$ are plotted for both species $s=1$ and $s=2$. These results were obtained by averaging over $280$ independent runs with different initial conditions. The perturbation was applied using $k={2 \pi \over L_{z}}$, where $L_{z}$ was the length of the simulation cell along the Cartesian $z-$direction. For the perturbation strength, the value $b=0.30$ was used. The response shown in Fig. \ref{ksq} most clearly corresponds to damped compression waves. As would be expected, the smaller mass for species 2 results in larger thermal fluctuations.

\begin{figure}
\begin{centering}
\includegraphics[width=0.5\textwidth]{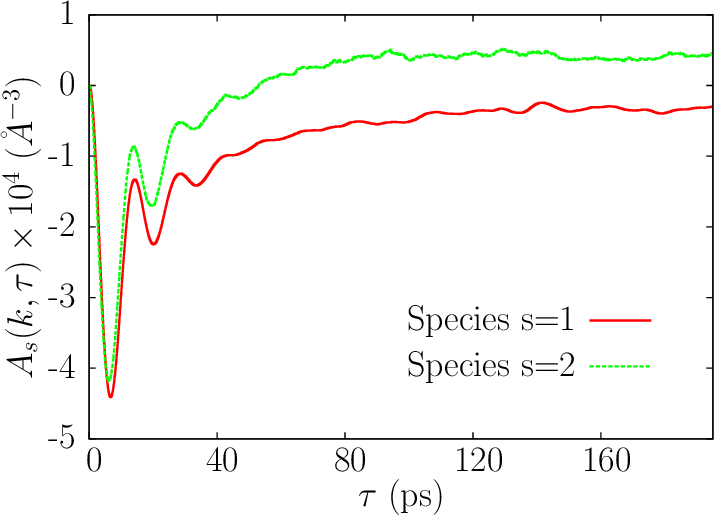} 
\caption{The function $\tilde{A}_{s}(k,\tau)$ obtained for an external perturbation at time zero for the system plotted for both species $s=1$ and $s=2$ with mass ratio ${m_{1} \over m_{2}}=8$.  The perturbation and response correspond to $k={2 \pi \over L}$.  The parameter $b=0.30$ for the perturbation strength was used.
}
\label{As}
\end{centering}
\end{figure}

While the data in Fig. \ref{ksq} demonstrates different responses for the two species, time-integration to obtain the density profiles more clearly shows the Soret effect.
Analysis of the particle distributions computed using Eq. \ref{dist1} are shown in Fig. \ref{As} from the same simulation data shown in Fig. \ref{ksq}. Specifically, the function $A_{s}(k,\tau)$ is shown for both components $s=1$ and $s=2$. The results show  that the pressure gradient created by the heat perturbation causes mass to flow away from the hot region, indicated by negative values for $A_{s}(k,\tau)$, followed by a series of damped oscillations. In comparison to the oscillation period, heat conduction occurs over a longer timescale.  As heat is conducted in the approach to thermal equilibrium, both species flow back to equilibrate the density. Fig. \ref{As} shows that during this process the concentration of species 2 (low-mass atoms) is enriched in the high-temperature region, and species 1 (high-mass atoms) is enriched in the low-temperature region. This is consistent with previous observations in binary mixtures\cite{Reith_2000}.

The results in Fig. \ref{As} provide insight into the physical mechanism  responsible for the Soret effect. During the evolution system after a heat pulse, mass flows occur in response to pressure gradients. Initially, the pressure gradient is caused by the heat pulse, but as density gradients develop in the system, the pressure gradients tend to diminish. This approach to near mechanical equilibrium includes compressional waves which are gradually damped.  Importantly, the concentration gradient develops during the period where compressional waves are pronounced. Over long timescales, heat conduction gradually restores thermal equilibrium, and the mass currents act to reduce the density gradient in the system.
In the case of the isotopic Soret effect shown in Figs. \ref{ksq}-\ref{As}, the only difference between the two species is their relative masses. It can be seen in Fig. \ref{As} that the high-mass species tends to overshoot during the initial flow away from the high-temperature region, and then over longer times responds more sluggishly as atoms flow back during the return to equilibrium. With each subsequent oscillation period, the density gradient is diminished, but the concentration gradient actually grows. This process leaves the high-temperature region enriched in low-mass atoms, and the low-temperature region enriched in high-mass atoms.

\begin{figure}
\begin{centering}
\includegraphics[width=0.5\textwidth]{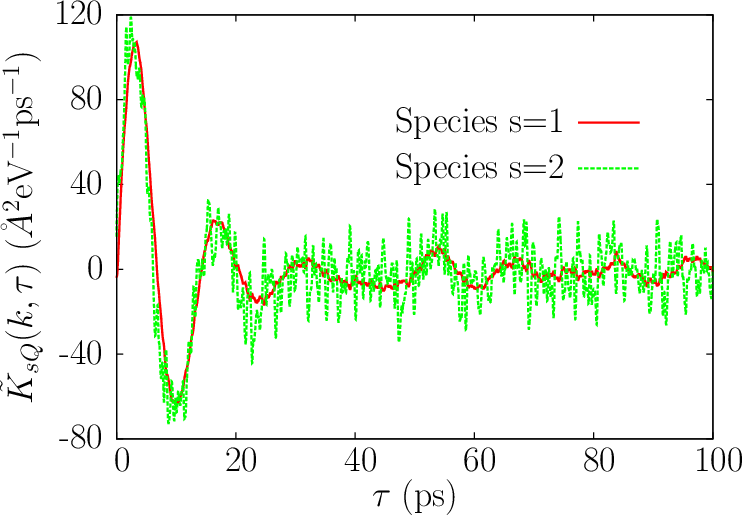} 
\caption{Response functions $\tilde{K}_{sQ}(k,\tau)$ obtained for an external perturbation at time zero for the system plotted for both species $s=1$ and $s=2$ with mass ratio ${m_{1} \over m_{2}}=8$. The perturbation and response correspond to $k={2 \pi \over L}$. Result obtained for perturbation strength $b=0.10$.
}
\label{ksq2}
\end{centering}
\end{figure}

\begin{figure}
\begin{centering}
\includegraphics[width=0.5\textwidth]{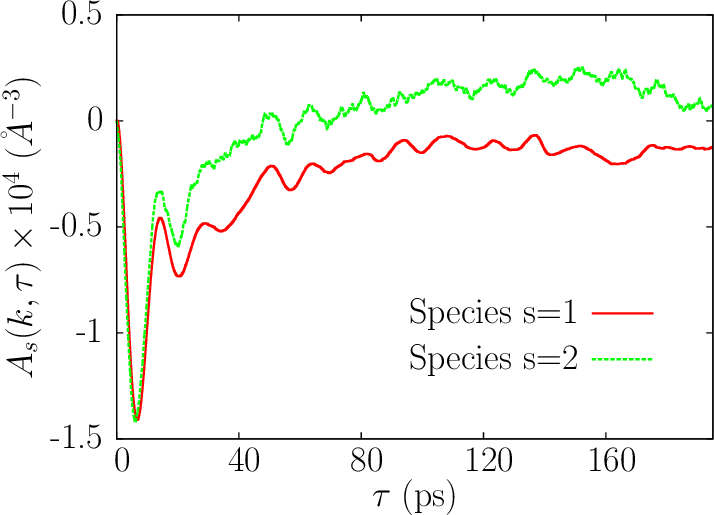} 
\caption{The function $A_{s}(k,\tau)$ obtained for an external perturbation at time zero for the system plotted for both species $s=1$ and $s=2$ with mass ratio ${m_{1} \over m_{2}}=8$. The perturbation and response correspond to $k={2 \pi \over L}$. Results obtained for perturbation strength $b=0.10$.
}
\label{As2}
\end{centering}
\end{figure}

Before moving on to other systems, it is important to demonstrate what range of values for the perturbation strength $b$ result in linear response. To establish a range of values, the calculations for the isotopic Soret effect were repeated with the smaller perturbation strength $b=0.10$. The response functions computed using $b=0.10$ are shown in Fig. \ref{ksq2}.  Comparison with Fig. \ref{ksq} shows that the results are essentially identical. The only noticeable difference is the presence of a relatively smaller signal-to-noise ratio in Fig. \ref{ksq2} which is to be expected. Integration of the results to obtain $A_{s}(k,\tau)$ tends to reduce the significance of the thermal noise as seen in Fig. \ref{As2}. Finally, the magnitude of the function $A_{s}(k,\tau)$ appears to depend linearly on the perturbation strength $b$, as seen by comparing Fig. \ref{As} and Fig. \ref{As2}. These results are expected if simulation with both $b=0.10$ and $b=0.30$ lie within the linear-response regime. Having established a range for linear response, the remaining calculations to be presented were performed using values generally less than $b=0.30$.

\begin{figure}
\begin{centering}
\includegraphics[width=0.5\textwidth]{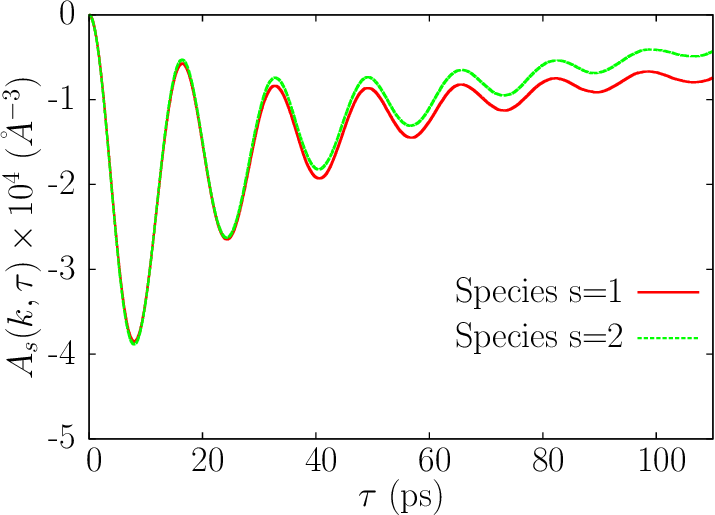} 
\caption{The function $A_{s}(k,\tau)$ obtained for an external perturbation at time zero for the system plotted for both species $s=1$ and $s=2$ with energy ratio ${\epsilon_{11} \over \epsilon_{22}}=2.5$. The vector for the perturbation and the response was $k={2 \pi \over L}$ and $b=0.20$ was used. Results were averaged from an ensemble of $336$ independent calculations.
}
\label{Aseps}
\end{centering}
\end{figure}

Now the effect of varying other parameters will be reported, starting with the relative interaction strength. For the first calculations, a larger system with $N_{1}=6000$, $N_{2}=6000$ was used. In addition, the system was longer along the direction parallel to the perturbation vector $\bm{k}$.  Specifically, the relative lengths $L_{z}=3L_{x}=3L_{y}$ were used. The particle density and reduced temperature were the same as those used to compute the reference state for Table \ref{table1}.
The ratio ${\epsilon_{11} \over \epsilon_{22}}=2.5$ was simulated with $m_{1}=m_{2}$ and $\sigma_{11}=\sigma_{22}$. In Fig. \ref{Aseps} the resulting function $A_{s}(k,\tau)$ is shown with the caption giving some of the simulation conditions. The general result is qualitatively similar to the results in Fig. \ref{As}, although the separation of the two components is somewhat less dramatic in comparison. In agreement with previous computational results\cite{Reith_2000}, Fig. \ref{Aseps} demonstrates that component $s=2$ tends to accumulate in the hot region, and component $s=1$ accumulates in the cold region. As with the isotope effect, the concentration gradient develops immediately during the first oscillation period of the compressional wave, and tends to grow with each subsequent oscillation of the liquid.

\begin{figure}
\begin{centering}
\includegraphics[width=0.5\textwidth]{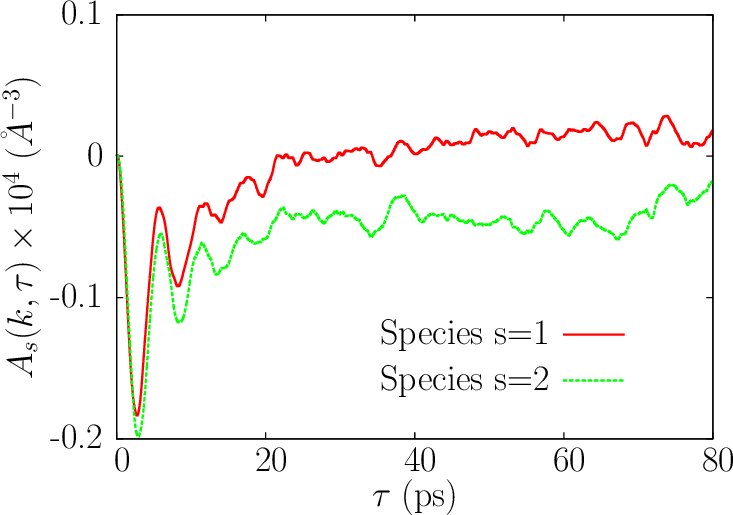} 
\caption{The function $A_{s}(k,\tau)$ obtained for an external perturbation at time zero for the system plotted for both species $s=1$ and $s=2$ with size ratio ${\sigma_{11} \over \sigma_{22}}=1.9$. The vector for the perturbation and the response was $k={2 \pi \over L}$ and $b=0.10$ was used. Results were averaged from an ensemble of $280$ independent calculations.
}
\label{Assig}
\end{centering}
\end{figure}

The effect of varying particle size is explored next. Simulations with ${\sigma_{11} \over \sigma_{22}}=1.9$ were performed, with the results for  $A_{s}(k,\tau)$ shown in Fig. \ref{Assig} with simulation details shown in the caption. The qualitative behavior is the same as in Fig. \ref{As} and Fig. \ref{Aseps}. The results show that the larger particles, corresponding to component $s=1$, tend to more readily return to the high-temperature region. This result is also in qualitative agreement with previous published results\cite{Reith_2000}, which demonstrated the enrichment of large particles in the high-temperature region and small particles in the low-temperature region. As with the other cases reported, the concentration gradient develops almost immediately, with a timescale comparable to the period of the compressional waves.

In summary, the method of generating a heating pulse results in rapid separation of two species in a Lennard-Jones fluid. This is in clear qualitative agreement with recent reports\cite{Hafskjold_2017}. Moreover, the effect of varying mass, interaction strength, and particle size separately was demonstrated to be in qualitative agreement with past results\cite{Reith_2000}.  In the next section, the mechanism causing the concentration gradients is explained. The qualitative description of the Soret effect clearly depends on a near balance between temperature gradients and density gradients which allow evolution in a state near mechanical equilibrium. It should also be noted that in contrast to the heat-pulse simulations here, the previous results were obtained using steady-state temperature gradients\cite{Reith_2000}. The next section will also establish the connection between the response-function and steady-state simulations.

\section{Analysis}

Here we analyze the particle current response to heat pulses. We first assume that currents act along only one coordinate direction. This corresponds exactly to the perturbation calculations presented earlier. Starting with the definitions of the reciprocal-space current terms $\tilde{J}_{1}$ and $\tilde{J}_{2}$ defined by Eq. \ref{js}, we consider changes that occur over a small time $\delta t$. Taking a time derivative we obtain,
\begin{equation} \label{jsexp}
\delta \tilde{J}_{s}(k,t)= {1 \over \Omega} \sum_{j=1}^{N_{s}}{1 \over m_{s}}\left[ F_{sj,z} - ikm_{s}v^{2}_{sj,z}\right]e^{-ikz_{sj}} \delta t ,
\end{equation}
in which the summation is conducted only over the $N_{s}$ atoms of species $s$.
The first term on the right-hand side includes the force component $F_{sj,z}$ acting on particle $n$ of species $s$ due to the interactions. The subscript $z$ indicates the Cartesian component associated with the wave vector $\vec{k}$ of the perturbation. The second term on the right-hand side depends on the velocity component $v_{sj,z}$ for each particle $j$ of species $s$, and is related to fluid flow resulting from a nonuniform distribution of kinetic energy. Considering these terms, we define the force term acting on species $s$ in reciprocal space,
\begin{equation} \label{force}
\tilde{F}_{s}(k)=\sum_{j=1}^{N_{s}} F_{sj,z}e^{-ikz_{sj}} .
\end{equation}
which only describes the forces along the direction of the perturbation.
For the term in Eq. \ref{jsexp} related to the kinetic energy of the particles, a local energy density $E_{s}(z)$ is defined for both species,
\begin{equation} \label{esr}
E_{s}(z)={1 \over A}\sum_{j=1}^{N_{s}}{1 \over 2}m_{s} v^{2}_{sj,z} 
\delta \left(z-z_{sj} \right) \text{     ,}
\end{equation}
in which $A=L_{x}L_{y}$ is the cross-sectional area of the simulation supercell.
This quantity can be written in reciprocal space as,
\begin{equation} \label{ke}
\tilde{E}_{s}(k) = {1 \over  AL_{z}} \sum_{n=1}^{N_{s}}{1 \over 2}m_{s} v^{2}_{sj,z} e^{-ikz_{sj}} \text{     .}
\end{equation}
Here $L_{z}$ is the length of the supercell along the direction parallel to the perturbation wave vector $\vec{k}$ and the system volume is $\Omega=AL_{z}$.
Then the evolution equations are,
\begin{equation} \label{jsk}
\delta \tilde{J}_{s}(k,t)= {1 \over m_{s} \Omega} \left[ \tilde{F}_{s}(k,t)-2ik \Omega \tilde{E}_{s}(k,t)\right] \delta t  \text{     .}
\end{equation}
This equation also appears in the Bearman-Kirkwood theory of transport in multicomponent systems\cite{Bearman_1958}, although in real rather than reciprocal space.

We next turn to the specific excitation source given by Eq. \ref{pertrecip} along with the current expression Eq. \ref{jsk} above. After integration given the particular source, it is then easy to show  that the response functions are,
\begin{equation} \label{respdetail}
\tilde{K}_{sQ}(k,t) = -{4 \over 3m_{s}Nbk_{B}T_{0}} \int_{0}^{t} \left[{1 \over k} \tilde{F}_{s}^{(i)}(k,t^{\prime})  - 2 \Omega \tilde{E}_{s}^{(r)}(k,t^{\prime}) \right] dt^{\prime} \text{     ,}
\end{equation}
in which $\tilde{F}_{s}^{(i)}=\Im \left[\tilde{F}_{s} \right]$ and $\tilde{E}_{s}^{(r)}=\Re\left[\tilde{E}_{s}\right]$. This equation has been directly tested by computing the terms $\tilde{F}_{s}(k,t)$ and $\tilde{E}_{s}(k,t)$ at each MD step after the perturbation and then integrating in time. The resulting response functions $\tilde{K}_{sQ}(k,t)$ obtained in this way were demonstrated to be identical to those obtained using Eq. \ref{respcalc}.

\begin{figure}
\begin{centering}
\includegraphics[width=0.5\textwidth]{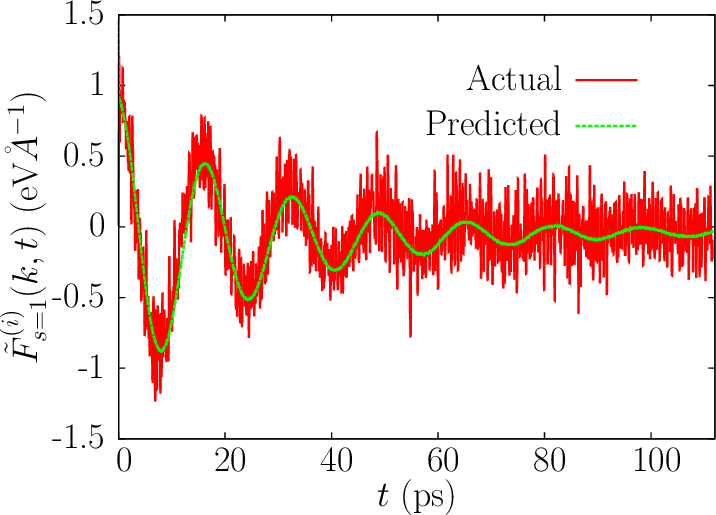} 
\caption{Imaginary part of the force term $\tilde{F}^{(i)}_{s}(k,t)$ from Eq. \ref{force} for species $s=1$ (solid red line). Calculations performed for a system with $N_{1}=6000$ and $N_{2}=6000$, and ${\epsilon_{11} \over \epsilon_{22}}=2.5$. The ensemble averaging was performed for $210$ independent initial conditions. Comparison is made to the theoretical expression Eq. \ref{forcek} (dashed green line).
}
\label{f1}
\end{centering}
\end{figure}

\begin{figure}
\begin{centering}
\includegraphics[width=0.5\textwidth]{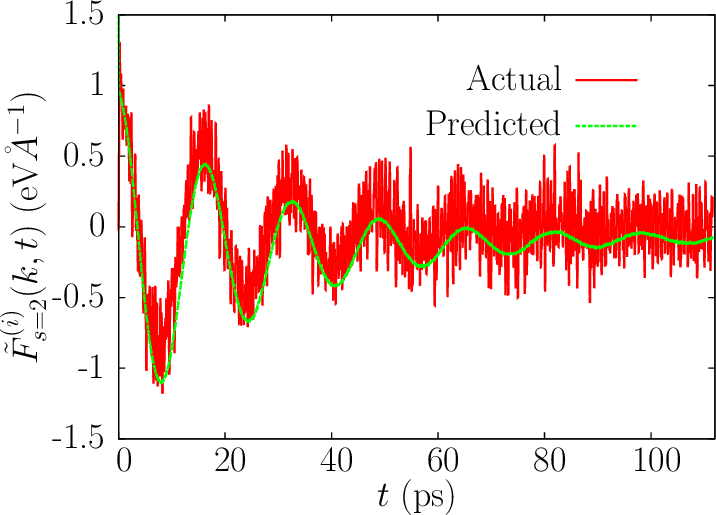} 
\caption{Imaginary part of the force term $\tilde{F}^{(i)}_{s}(k,t)$ from Eq. \ref{force} for species $s=2$ (solid red line). Calculations performed for a system with $N_{1}=6000$ and $N_{2}=6000$, and ${\epsilon_{11} \over \epsilon_{22}}=2.5$. The ensemble averaging was performed for $210$ independent initial conditions. Comparison is made to the theoretical expression Eq. \ref{forcek} (dashed green line).
}
\label{f2}
\end{centering}
\end{figure}

The equation above shows how the currents evolve due the force term $\tilde{F}_{s}(k,t)$. The insight obtained previously suggests that the Soret effect, which depends on evolution in near mechanical equilibrium conditions, can be explained by understanding how $\tilde{F}_{s}(k,t)$ depends on temperature and concentration gradients. In the Bearman-Kirkwood theory \cite{Bearman_1958}, which was the first attempt at a statistical mechanics representation of transport theory in multicomponent systems, the driving force for the Soret effect arises due to temperature gradients. Specifically, it was proposed by Bearman and Kirkwood\cite{Bearman_1958} and used later by Morozov \cite{Morozov:2009wa} that the driving force for species $s$ is,
\begin{equation}\label{morozov}
\bm{F}_{s} = -\left({\partial p_{s} \Omega \over \partial T} \right)_{p,w} \bm{\nabla}T \text{     ,}
\end{equation}
which is to be taken at fixed values of the weight fraction $w$ and pressure $p$. This implies a driving force only due to a temperature gradient, and in fact implies mechanical equilibrium. However, in the current calculations, the system responds to a temperature perturbation and is explicitly not in mechanical equilibrium as it evolves. Moreover, as the simulations demonstrate, the forces depend on not just the temperature gradient, but also on the density gradient that is established. Hence, Eq. \ref{morozov} only represents part of the driving force responsible for evolution. For our purposes, we hypothesize that the driving force can be written in reciprocal space,
\begin{equation} \label{forcek}
\tilde{F}_{s}(k)=-ik \Omega\left[
\left( {\partial p_{s} \over \partial T}\right)_{n_{1},n_{2}} \tilde{T}(k)+ \left( {\partial p_{s} \over \partial n_{1}}\right)_{T,n_{2}} \tilde{n}_{1}(k)+ \left({\partial p_{s} \over \partial n_{2}} \right)_{T,n_{1}}\tilde{n}_{2}(k) 
\right] \text{     ,}
\end{equation}
in which $\tilde{T}(k)$, defined below, is related to temperature gradients, and $\tilde{n}_{s}(k)$ represent density gradients in reciprocal space for both species $s=1,2$. This equation accounts for the fact that temperature and density gradients generate forces and consequently currents. It accounts for the observation that heat pulses lead to density gradients, with mechanical equilibrium corresponding to the point when $\tilde{F}_{s}(k)$ tends to zero for both species, corresponding to low density in the high-temperature region and high density in the low-temperature region.
It should be noted that the calculations include an ensemble of systems prepared in distinct equilibrium before the heat perturbation. Hence, each quantity in Eq. \ref{forcek} will represent an ensemble-averaged quantity. Also, it should be understood that dissipative frictional forces exist\cite{Hoang_2013}, for example due to relative motion of the two components. These are neglected in the above expression Eq. \ref{forcek}. However, for our purposes here, Eq. \ref{forcek} provides a quantity that demonstrates competing effects due to temperature and density gradients, and furthermore provides insight into both the origin of the Soret effect and the sign of the Soret coefficient. 

To use Eq. \ref{forcek}, a definition for the approximate local temperature is required. Here,
the expression for the local temperature $T(z)$ in is defined using the local kinetic-energy density $E_{K}(z)$ and particle density $n(z)=n_{1}(z)+n_{2}(z)$. Applying the classical equipartition theory, 
\begin{equation}
{E_{K}(z) \over  n(z)}= {3 k_{B} \over 2} T(z) \text{     .}
\end{equation}
The expression for $E_{K}(z)$ follows the general form for $E_{s}(z)$ in Eq. \ref{esr}, but includes both species and also velocity components in the $x-$ and $y-$Cartesian directions, specifically,
\begin{equation}
E_{K}(z)={1 \over A} \sum_{s=1}^{2} \sum_{j=1}^{N_{s}} {1 \over 2}m_{s}(\bm{v}_{sj}\cdot \bm{v}_{sj}) \delta(z-z_{sj}) \text{    .}
\end{equation}
This is represented in reciprocal space by $ \tilde{E}_{K}(k)$ using,
\begin{equation}
\tilde{E}_{K}(k)={1 \over AL_{z}} \sum_{s=1}^{2} \sum_{j=1}^{N_{s}} {1 \over 2}m_{s}(\bm{v}_{sj}\cdot \bm{v}_{sj}) e^{-ikz_{sj}} \text{     .} 
\end{equation}
Likewise, the particle density is represented in reciprocal space by $\tilde{n}(k)$ using,
\begin{equation}
\tilde{n}(k)={1 \over AL_{z}}  \sum_{s=1}^{2} \sum_{j=1}^{N_{s}} e^{-ikz_{sj}} \text{     .}
\end{equation}
Then the temperature in reciprocal space can represented approximately as,
\begin{equation} \label{tk}
\tilde{T}(k) = T_{0}\left[{2 \tilde{E}_{K}(k)\over 3 n_{0} k_{B} T_{0}} - {\tilde{n}(k) \over n_{0}} \right] \text{     ,}
\end{equation}
in which $n_{0} = {N_{1}+N_{2} \over \Omega}={N \over \Omega}$. In obtaining Eq. \ref{tk}, it was assumed that perturbations are relatively small, and terms quadratic in any reciprocal space term for finite $k$ was neglected. There is another assumption that current flows are relatively small, such that the kinetic energy associated with local current flows can be included in the definition of $\tilde{T}(k)$ without large errors. 

The force term $\tilde{F}_{s}(k,t)$ from Eq. \ref{force} was computed after an excitation during the simulations of the ${\epsilon_{11} \over \epsilon_{22}}=2.5$ system. To make a comparison to Eq. \ref{forcek}, the ensemble-averaged temperature $\tilde{T}(k)$ and densities $\tilde{n}_{s}(k)$ were computed in reciprocal space. As previously, the excitation corresponded to $k={2 \pi \over L_{z}}$. 

The results for species $s=1$ and $s=2$ are shown in Fig. \ref{f1} and Fig. \ref{f2}  respectively. Comparison of Fig. \ref{f1} and Fig. \ref{f2} show no obvious differences. The oscillatory behavior noted previously in the response functions is also evident in the forces. This arises due to compression waves which propagate through the system as has been describe above. The calculated forces using Eq. \ref{forcek} are also presented in Figs. \ref{f1}-\ref{f2} showing overall qualitative agreement. This demonstrates that the prediction based on Eq. \ref{forcek} is entirely reasonable, and further shows the competing effects due to pressure and density gradients.

While the theoretical expression in Eq. \ref{forcek} for $\tilde{F}_{s}(k,t)$ is only semi-quantitative, it does provide insight into the mechanism of the Soret effect. Specifically, the Soret effect occurs due to the fact that heat pulses generate pressure gradients which drive particle currents. Initially, the heat pulse generates a large pressure gradient which drives particles away from the high-temperature region. Subsequent oscillations persist but gradually dissipate.  
The Soret effect can be understood then as arising due to differences in how the two species in a binary liquid respond to the evolving pressure gradients. As seen previously, the isotope effect occurs due to the fact that the light species responds more rapidly to evolving conditions. Similar considerations were first made by Galliero and coworkers\cite{GALLIERO2003171}. For the other cases simulated here, the data in Table \ref{table1} is key to understanding the Soret effect. First, it is evident that the species which accumulates in the high-temperature region also corresponds to the species with the largest partial pressure. In the case of simulations with ${\epsilon_{11} \over \epsilon_{22}}=2.5$, species $s=2$ has the largest partial pressure ($p_{2} > p_{1}$) and also is the species which accumulates in the high-temperature region. Similarly, for simulations with ${\sigma_{11} \over \sigma_{22}}=1.9$, the large particles corresponding to species $s=1$ have the largest partial pressure ($p_{1} > p_{2}$) and also tend to accumulate in the high-temperature region.

However, the Soret effect is most clearly understood when considering the relative sensitivity of the partial pressures $p_{s}$ to changes in temperature and density. These are also given in Table \ref{table1}. First, the positive values for $\left({\partial p_{s} \over \partial T}\right)$ are consistent with the main effect of the heat pulse, which is to drive both species away from the high-temperature region. However, as particle densities increase in the low-temperature region, the density gradient results in an opposing pressure gradient. This dynamic leads to the compressional waves. The data in Table \ref{table1} shows that the species with the largest partial pressure $p_{s}$ is also more sensitive to pressure and density gradients. What is key in considering the relative sensitivities is that the temperature derivatives of $p_{s}$ differ less between the two species than concentration derivatives. Hence, if one considers the simulations with ${\sigma_{11}\over \sigma_{22}}=1.9$, the initial temperature gradient due to the heat pulse is expected to drive species $s=1$, the large particles, more strongly away from the high-temperature region. However, the data in Table \ref{table1} suggest that the most important effect is due to the density gradient, with species $s=1$ driven away from regions of high particle density more strongly than species $s=2$. The same considerations can be applied to simulations with ${\epsilon_{11}\over \epsilon_{22}}=2.5$, with in this case the dominant effect is that species $s=2$ will be strongly driven away from the high-density region.

These considerations  provide a clear qualitative picture to interpret the results for $A_{s}(k,t)$ shown in Figs. \ref{Aseps}-\ref{Assig}. The temperature gradient imposed by the heat pulse generates a pressure gradient for both species. This contribution to the pressure gradient always acts to push particles away from the high-temperature region, with the effect largest for the species $s$ with the largest value for ${\partial p_{s} \over \partial T}$. This effect, however, is more than compensated by the density gradient. The density gradient contributes to the pressure gradient in a direction which always acts to push particles away from the high-density (low-temperature) region back to the low-density (high-temperature) region,  with the largest effect for the species with the largest values for ${\partial p_{s} \over  \partial n_{s^{\prime}}}$. This latter effect is dominant and is responsible for creating the concentration gradient. Upon each mechanical oscillation of the system, the concentration gradient is found to increase in Figs. \ref{Aseps}-\ref{Assig}.

\begin{figure} 
\begin{centering} 
\includegraphics[width=0.5\textwidth]{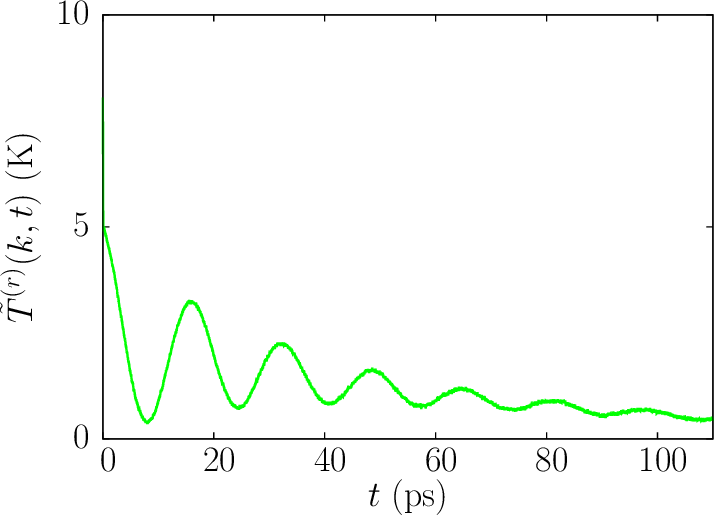} 
\caption{Real part of the computed reciprocal-space temperature $\tilde{T}^{(r)}(k,t)$ for $k={2\pi \over L_{z}}$. The plot shows oscillations due to compression waves, but also the very gradual return to thermal equilibrium which tends to drive the separation of the two species in the liquid.
}
\label{tempk}
\end{centering}
\end{figure}

During the evolution back to equilibrium, heat conduction occurs to return the system to thermal equilibrium. 
In contrast to the process of generating density and concentration gradient, the timescale for heat conduction is quite long. The process of attaining thermal equilibrium is seen in the decrease with time $t$ of the Fourier component $\tilde{T}(k,t)$. The time-dependence of the real component $\tilde{T}^{(r)}(k,t) = \Re{\left[ \tilde{T}(k,t)\right]} $ is shown in Fig. \ref{tempk} for the simulation ensemble with ${\epsilon_{11} \over \epsilon_{22}}=2.5$. Initially there is a very large jump corresponding to values $\tilde{T}^{(r)}(k,t) \approx 18$K representing the tendency of the heat pulse to rapidly drive atoms from the high-temperature region to the low-temperature region. Due to the oscillations, these temperatures  likely do not represent accurately the local temperature. However, after the oscillations dissipate,  $\tilde{T}(k,t)$ becomes a more precise measure of the departure from thermal equilibrium. Over longer times, the values for $\tilde{T}^{(r)}(k,t)$ are not too different from the predicted value $\tilde{T}^{(r)}(k,t)\approx 6.5$K based on the average temperature $T_{0}=161$K and the value $b=0.20$ used for the heat pulse.  Finally, given that $T_{0}=161$K for the equilibrium temperature, the result in Fig. \ref{tempk}, which exhibit relatively small deviations from equilibrium, are consistent with the contention that the system is perturbed within the regime where linear response is expected.

It is important finally to make a connection to the Soret effect in steady-state  conditions, in which the Soret coefficient can be determined based on calculation of the concentration and temperature gradients. Here it can be noted that 
the response functions $\tilde{K}_{sQ}(k,t)$ represent Green's functions for heat perturbations. Consequently, if these functions are accurately known, they can be used for the description of particle currents that arise due to a time-dependent perturbations, possibly with different spatial characteristics. Here the effect of a time-independent heating power $\tilde{H}^{(ext)}(k)$ will be established both by direct simulation and using the response function $\tilde{K}_{sQ}(k,t)$. This will determine the concentration gradient in the presence of a stationary heat current. Comparison to steady-state simulations was done to connect the response functions to the calculations reported in Ref.\cite{Reith_2000}.

\begin{figure} 
\begin{centering} 
\includegraphics[width=0.5\textwidth]{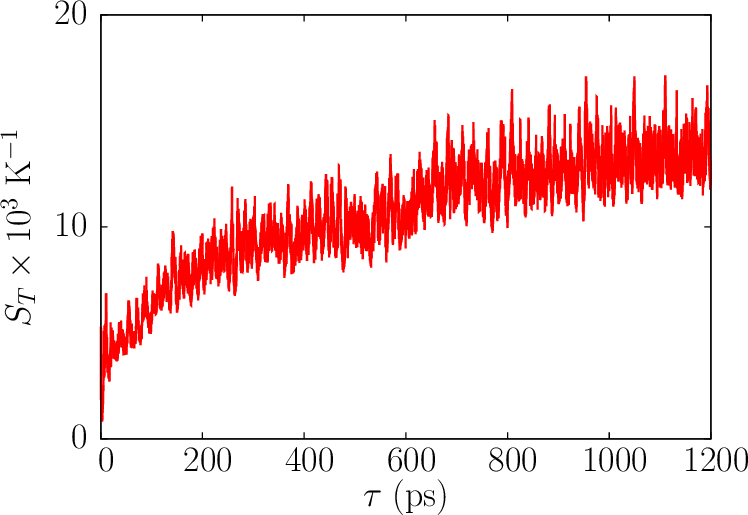} 
\caption{Calculation of the Soret coefficient $S_{T}$ as a function of simulation time. The system corresponded to the $N=4000$ particle system with ${m_{1} \over m_{2}}=8$. The heating was done using a sinusoidal heat pulse at each MD step with $b=2 \times 10^{-5}$ and $k={2 \pi \over L_{z}}$.
}
\label{soret}
\end{centering}
\end{figure}

Specifically, simulations with ${m_{1} \over m_{2}}=8$ and the same conditions reported above for this system were performed, with the addition of a steady-state sinusoidal heat perturbation.  In these calculations, the heat perturbation was applied at each MD step with a small perturbation strength $b=2 \times 10^{-5}$. The external heat power is then given by,
\begin{equation}
\tilde{H}^{(ext)}(k) = {3N b k_{B}T \over 4 \Omega \Delta t} \text{     ,}
\end{equation}
in which $\Delta t=6.086$fs is the MD timestep.  The Soret coefficient $S_{T}(\tau)$ is computed at each time $t$ from the reciprocal-space density $\tilde{n}_{1}(k,\tau)$ and kinetic energy $\tilde{E}_{K}(k,\tau)$ using the expression\cite{Reith_2000},
\begin{equation}
S_{T}(\tau) \approx -{6 k_{B} \tilde{n}_{1}(k,\tau) \over \tilde{E}_{K}(k,\tau)} \text{     ,}
\end{equation}
in which the kinetic energy term $\tilde{E}_{K}(k,\tau)$ is directly related to the temperature via the classical-equipartition theorem.
Although this quantity is generally defined in steady-state conditions, here it is plotted in Fig. \ref{soret} as a function of time to show the approach to steady state. The value obtained at $\tau=1.2$ns is about $S_{T} \approx 14 \times 10^{-3}$K$^{-1}$, which is somewhat smaller than the value $S_{T}=24.6 \pm 0.8 \times 10^{-3}$K$^{-1}$ reported previously\cite{Reith_2000}. It is not certain why these results differ slightly, but it should be noted that the previous results were for different system sizes and also did not use sinusoidal heating profiles\cite{Reith_2000}. It is also possible that a longer calculation might result in a larger value for $S_{T}$. Given those considerations, the results are in relatively close agreement. While there are other results for LJ fluids in the literature\cite{SIMON1998151,Perronace:2002aa,Galli_ro_2003,YEGANEGI2005171}, they appear to all represent different sets of parameters and state points. However, it would be relevant and important to compare results obtained using other state points and simulation methods. Finally, the computed temperature difference between the high- and low-temperature regions at the end of the calculation was about $\sim 16$K, which is significantly less than the average temperature $102$K. 

\begin{figure} 
\begin{centering} 
\includegraphics[width=0.5\textwidth]{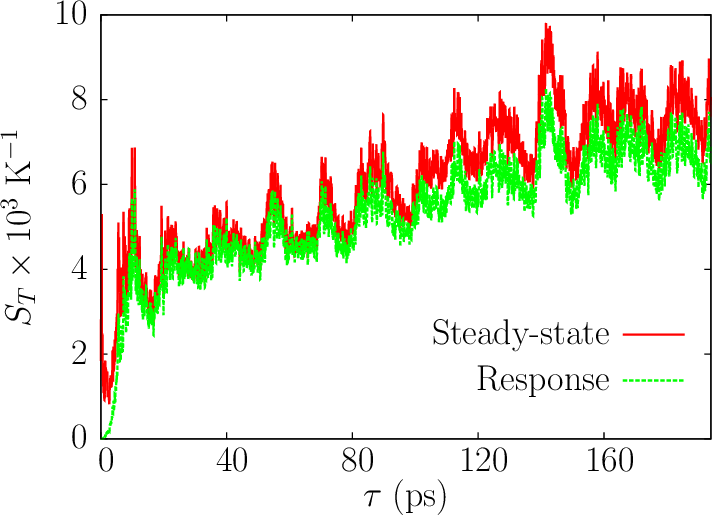} 
\caption{Calculation of the Soret coefficient $S_{T}$ using the data shown in Fig. \ref{soret} (labeled here ``Steady-state''), but here compared to the prediction based on the computed response function $\tilde{K}(k,t)$ from Eq. \ref{resppred} (labeled here ``Response'') . The response function was computed from an ensemble of $280$ simulations with independent initial conditions. The system corresponded to the $N=4000$ particle system with ${m_{1} \over m_{2}}=8$. The heating was done using a sinusoidal heat pulse at each MD step with $b=2 \times 10^{-5}$ and $k={2 \pi \over L_{z}}$.
}
\label{soretcomp}
\end{centering}
\end{figure}

Next the connection to the response functions and the results in Fig. \ref{soret} will be established.  First, an ensemble of longer  calculations were performed to obtain the response function $\tilde{K}(k,t)$ to  a maximum time $t=194$ps.  Given the reciprocal space heating power $\tilde{H}^{(ext)}(\vec{k})$ as defined above, the reciprocal-space particle currents are given by, 
\begin{equation}
\tilde{J}_{s} (k,\tau)=  -ik \tilde{H}^{(ext)}(k) \int_{0}^{\tau} \tilde{K}_{sQ}(k,t) dt \text{     ,}
\end{equation}
in which the initial ensemble represents equilibrium so that $\tilde{J}_{s} (k,0)=  0$.
Then using the continuity equation in reciprocal space and assuming that for an ensemble of initial states which reflect equilibrium by $\tilde{n}_{1}(k,0) =  \tilde{n}_{2}(k,0) =0$, the ensemble averages for $\tilde{n}_{s}(k,\tau)$ should be determined by,
\begin{equation}
\tilde{n}_{s}(k,\tau) = -k^{2} \tilde{H}^{(ext)}(k)  
\int_{0}^{\tau}d\tau^{\prime} \left[\int_{0}^{\tau^{\prime}} \tilde{K}_{sQ}(k,t) dt \right] \text{     .}
\label{resppred}
\end{equation}
Because the response functions were only obtained to a time $\tau=194$ps, it is only possible to compare with the simulation data shown in Fig. \ref{soret} to this time. It was found in fact that the response function after $194$ps was still consistent with a concentration gradient, and hence diffusion to return the system  to equilibrium without any concentration gradient would require an even longer calculation. Nevertheless, it is still possible to directly compare the predictions based on the response function used in Eq. \ref{resppred} with the data shown in Fig.\ref{soret}. This is shown in Fig. \ref{soretcomp}. Some interesting features emerge. As expected, since after $194$ps the concentration gradient is still increasing, the value of $S_{T}(\tau)$ is only about half the value at $\tau=1.2$ns. However, comparison of the two different approaches demonstrates excellent agreement, establishing the applicability of the response functions to determine $S_{T}$. Moreover, while the concentration gradient develops rapidly, the continuous heat pulse gradually builds the concentration gradient to larger values over longer times. In summary, these results validate the use of response functions to elucidate behavior in a steady-state calculation like those reported previously\cite{Reith_2000}.

\begin{figure} 
\begin{centering} 
\includegraphics[width=0.5\textwidth]{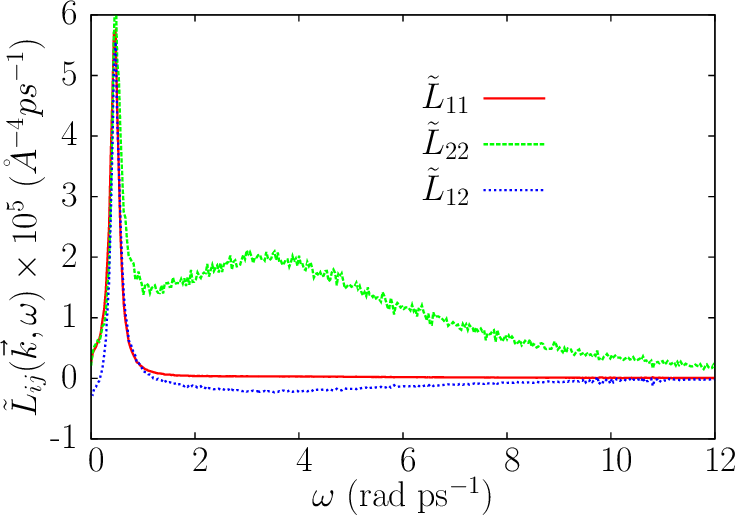} 
\caption{The Fourier transform of the correlation functions in Eq. \ref{onsager} for a system with ${m_{1} \over m_{2}}=8$. For this system the results show clearly the presence of ``fast sound'' modes involving motion of the low-mass atom of species $s=2$.
}
\label{jj1}
\end{centering}
\end{figure}

\begin{figure} 
\begin{centering} 
\includegraphics[width=0.5\textwidth]{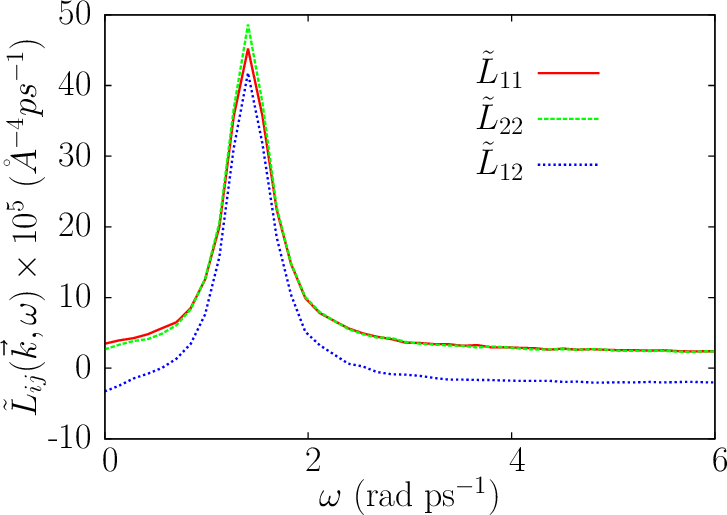} 
\caption{The Fourier transform of the correlation functions in Eq. \ref{onsager} for a system with ${\epsilon_{1} \over \epsilon_{2}}=2.5$.
}
\label{jj2}
\end{centering}
\end{figure}

\begin{figure} 
\begin{centering} 
\includegraphics[width=0.5\textwidth]{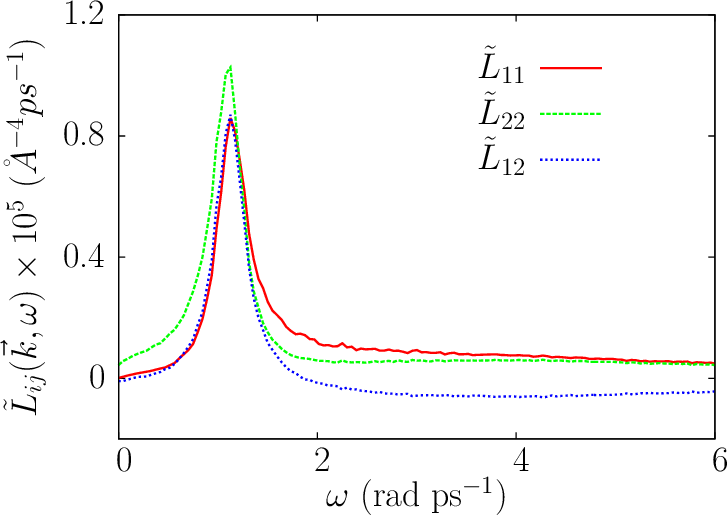} 
\caption{The Fourier transform of the correlation functions in Eq. \ref{onsager} for a system with ${\sigma_{1} \over \sigma_{2}}=1.9$.
}
\label{jj3}
\end{centering}
\end{figure}

The results above demonstrate that the Soret effect is essentially a mechanical effect. To further explore this mechanism, the reciprocal space correlation functions $\tilde{L}_{ij}(k,\tau)$ defined by,
\begin{equation}
\tilde{L}_{ij}(k,\tau) = \int_{0}^{\tau} \langle \tilde{J}_{i}(k,t)\tilde{J}_{j}(-k,t) \rangle dt \text{     ,}
\label{onsager}
\end{equation}
were computed. In Figs.\ref{jj1}-\ref{jj3} ensemble averages of the Fourier transforms of these function are shown for each system considered. Consistent with the propagation of compressional waves, each case exhibits a relatively clear peak but with substantial broadening. For the simulations with ${m_{1} \over m_{2}}=8$, the results show the clear existence of a low-frequency mode and a high-frequency mode. In the high-frequency mode, only the low-mass atoms $s=2$ participate. This mode corresponds to what is generally termed ``fast sound'' in the literature\cite{Bosse:1986vu,Schram_1990}, and clearly involves interdiffusion of the two species. For the other two cases with ${\epsilon_{11} \over \epsilon_{22}}=2.5$ (Fig. \ref{jj2}) and 
${\sigma_{11} \over \sigma_{22}}=1.9$ (Fig. \ref{jj3}), the presence of low- and high-frequency modes is not evident. However, the results show clearly that in both cases species $s=1$ and $s=2$ should be expected to respond differently to pressure gradients. Specifically, both species should largely move in concert, but also exhibit some interdiffusion when compressional waves propagate. This expected behavior is consistent with the computed response functions.

\section{Conclusions}

The results here demonstrate that the physical picture for the Soret effect depends on how binary liquids respond to pressure gradients induced by external heat sources.
To the best of my knowledge, the mechanism identified here appears to not have been previously considered. It is generally assumed that the response to mechanical imbalances is very rapid compared to diffusional timescales\cite{Howard_1964}. While these considerations are certainly valid, when applied to the Soret effect in liquids, they appear to obscure the mechanism. In fact, the usual assumption appears to be that the rapid establishment of mechanical equilibrium places constraints on thermodynamic driving forces\cite{Howard_1964}. These constraints are known as Prigogine's theorem\cite{deGroot_1952}. By contrast, the results here demonstrate that concentration gradients develop on the same timescales as compressional waves, but generally take very long to reach their steady-state values in the presence of a continuous heating power.

In developing understanding of the effect, it was demonstrated that light atoms tend to respond more quickly to pressure gradients, thereby accumulating in the high-temperature region. This phenomenon appears to correspond with the presence of ``fast sound'' modes and their clear connection to interdiffusion. For the other cases simulated, differences in the sensitivity of the partial pressures to temperature and density perturbations determine the behavior. It was found specifically that differences in sensitivity to density gradients tend to be more significant than differences in the sensitivity to temperature gradients. This leads to enrichment of the species with the largest partial pressure in the high-temperature region, and enrichment of the species with the lowest partial pressure in the low-temperature region.

Previous studies appear to have observed the same basic phenomena, but perhaps have not characterized it as completely as is presented here. The first articles that appear to characterize transient responses both note the presence of compressional sound waves\cite{Hafskjold_2017,Bonella_2017} in response an applied heat source. More details consistent with the picture developed here appear in the article by Bonella and coworkers\cite{Bonella_2017}. In addition to the gradual evolution of the overall particle density in parallel with the density gradient, it was noted that compressional waves are due to competing effects resulting from the overall temperature and density gradients\cite{Bonella_2017}. However, the differences in how the two species respond to the gradients was not connected to the Soret effect itself\cite{Bonella_2017}. Nevertheless, the results in \cite{Bonella_2017}, in particular Figs. 5-6, show a perceptible concentration gradient over a timescale which is comparable to the period of compressional waves. In the somewhat earlier work by Ferrario and coworkers, concentration differences began to be noticeable on picosecond timescales\cite{Ferrario_2016}. However, consistent with the results presented here, the work in \cite{Bonella_2017,Ferrario_2016} demonstrates that the concentration gradients accumulate over nanosecond timescales.

It is worth focussing on the relevant timescales for the problem, and the connection to experiment. There are primarily three timescales, each of which depends on the system size or typical length $L$ associated with a heat source. The shortest scale is associated with compressional waves, $\tau_{s} \sim {L \over v_{s}}$, where $v_{s}$ is the sound velocity. The intermediate timescale is related to thermal conduction, $\tau_{cond} \sim {L^{2} \over \lambda_{T}}$, in which $\lambda_{T}$ is the thermal diffusivity. This scale gives the time for heat conduction to restore a system to thermal equilibrium after a heat pulse. The longest relevant timescale is that of ordinary interdiffusion, $\tau_{diff} \sim {L^{2} \over D_{12}}$, where $D_{12}$ is the coefficient for interdiffusion. This scale is related to the time required for a concentration gradient to relax back to equilibrium. From the response functions computed here, $\tau_{diff}$ can easily approach $\sim 1$ns for the simulation sizes reported here. Hence, while concentration gradients begin to form on very short time scales, the resulting concentration gradients are very slow to relax back to equilibrium, and a constant heating source tends to gradually build the concentration gradient. In experiment, therefore, the concentration gradient will generally be observed as something that builds very gradually, and the connection to mechanical oscillations might be challenging to observe directly. Moreover, since the diffusional timescales scale as $L^{2}$, whereas the timescale for oscillations scale as $L$, the separation in the scales used in typical experiments is likely to be much larger than what is characteristic of the simulations here, for which $L \sim 17$nm was used. Consequently, it might be possible to more easily demonstrate the transients or short time behavior via experiments with heat sources characterized by smaller $L$. In experiments using thermal diffusion forced Rayleigh scattering (TDFRS)\cite{kohler_book,Koehler_1995}, the diffraction length scale is typically $L \sim 10 \mu$m\cite{Perronace_2002}. Hence the experimental length scale is about $10^{3}$ times larger than the simulations reported here. In experiment then, diffusion timescales $\tau_{diff}$ of at least several milliseconds\cite{Debuschewitz:2001aa} are observed, which is greater by a factor $10^{6}$ than the diffusion timescales for the simulations. Consequently, the separation of scales is dramatically greater in experiment, and on timescales characteristic of compressional waves, the amount of separation is expected to be too small to measure.

It is interesting to speculate whether conditions might exist in which the species with the largest partial pressure accumulates instead in the low-temperature region.  The qualitative insight obtained in this paper suggests that this would occur if the partial pressures were more sensitive to temperature gradients rather than density gradients. This would correspond to ${\partial p_{s} \over \partial T}$ playing the role as the most significant difference between the two species.  It should be noted that both sensitivity to temperature and density gradients themselves must depend on the average density (or pressure) and temperature. This could potentially explain explain various results in the literature of the sign of $S_{T}$ changing as average conditions are varied\cite{Duhr_2006}. However, other possible explanations have been proposed\cite{Duhr_2006}.

It is also suggested  that the somewhat loosely defined single-component property ``thermophobicity'' might be associated with  the data in Table \ref{table1}. Specifically, thermophobicity might depend on the relative importance of partial derivatives with respect to temperature ${\partial p_{s} \over \partial T}$ and density  ${\partial p_{s} \over  \partial n_{s}^{\prime}}$.  This might suggest some directions to better quantify what exactly controls thermophobicity.

It is also interesting to contrast the approach developed here and the approach in Ref. \cite{Reith_2000} with Green-Kubo calculations. 
In the Green-Kubo approach, generally time-correlation functions are obtained for the total, hence $\vec{k}=0$, currents. For the mass currents, when periodic boundary conditions are applied, Newton's third law requires that,
\begin{equation} \label{thirdlaw}
m_{1} \bm{J}_{1}(t) + m_{2} \bm{J}_{2}(t)=0 \text{     ,}
\end{equation}
at all times $t$. This constraint is well-known to result in only one independent heat of transport\cite{Evteev_2014,TUCKER201654}, which is closely related to the Soret coefficient $S_{T}$. By contrast, the other computational approaches, while obeying Eq. \ref{thirdlaw} overall, do not require Eq. \ref{thirdlaw} to be obeyed locally. Hence, in the simulations reported here, both components flow away from the high-temperature region after the input heat pulse. In reciprocal space, this implies that for $\bm{k}\ne 0$, both $\tilde{J}_{1}(k,t)$ and $\tilde{J}_{2}(k,t)$ are allowed to vary independently without any constraint. This evolution after a heat pulse depends on the induced pressure gradient, which has been shown to be closely connected to the generation of the concentration gradient. In a Green-Kubo calculation, there can be no equilibrium pressure fluctuations corresponding to a pressure gradient with $\bm{k}=0$. It would be interesting to instead apply the Green-Kubo approach but considering fluctuations with $\bm{k} \ne 0$, which would allow equilibrium pressure fluctuations to be correlated with currents and concentration gradients. Also, performing Green-Kubo calculations for $\bm{k} \ne 0$ quantities would allow for two independent heats of transport in a two component system. More broadly speaking, there seems to be a lack of efforts to validate different simulation approaches with Green-Kubo in the case of the Soret effect. This appears to be another direction to explore.

Finally it should be recognized that the results presented here do not suggest any reason to revisit the phenomenological equations based on linear irreversible thermodynamics. In the standard picture, coupled equations for heat and mass transfer are used with Onsager transport coefficients describing the relations between driving forces and response currents. The objective of this paper is to identify the microscopic mechanism for the Soret effect, and to connect the Onsager transport coefficients to quantities that can be measured in equilibrium. None of the results obtained here would indicate any separate account of the separation that occurs on very fast time scales would need to be included explicitly within the usual linear irreversible thermodynamics framework.

\section{Acknowledgements} 
The calculations reported here were performed using the STOKES computing cluster at UCF. While this work did not have external funding, some of the ideas were developed over time based on initial work done on an NSF-funded project 1106219. I would also like to thank the referees, who provided very useful comments to improve the presentation of the work.

\newpage

\bibliographystyle{unsrt}

\begin{thebibliography}{10}

\bibitem{ludwig_1856}
C.~Ludwig.
\newblock {\em Akad. Wiss. Wien}, 20:539, 1856.

\bibitem{soret_1879}
C.~Soret.
\newblock {\em Arch. Geneve}, 3:48, 1879.

\bibitem{Morozov:2009wa}
Konstantin~I. Morozov.
\newblock Soret effect in molecular mixtures.
\newblock {\em Physical Review E}, 79(3):031204, 2009.

\bibitem{Reith_2000}
Dirk Reith and Florian M{\"u}ller-Plathe.
\newblock On the nature of thermal diffusion in binary Lennard-Jones liquids.
\newblock {\em The Journal of Chemical Physics}, 112(5):2436--2443, 2000.

\bibitem{Hoang_2022}
Hai Hoang and Guillaume Galliero.
\newblock Predicting thermodiffusion in simple binary fluid mixtures.
\newblock {\em The European Physical Journal E}, 45(5):42, 2022.

\bibitem{Hartmann:2012va}
S.~Hartmann, G.~Wittko, W.~K{\"o}hler,   K.~I. Morozov, K.~Albers, and G.~Sadowski
\newblock Thermophobicity of liquids: Heats of transport in mixtures as pure
  component properties.
\newblock {\em Physical Review Letters}, 109(6): 065901, 2012.

\bibitem{Hartmann_2014}
S.~Hartmann, G.~Wittko, F.~Schock, W.~Gro{\ss}, F.~Lindner, W.~K{\"o}hler, and
  K.~I. Morozov.
\newblock Thermophobicity of liquids: Heats of transport in mixtures as pure
  component properties{\textemdash}the case of arbitrary concentration.
\newblock {\em The Journal of Chemical Physics}, 141(13):134503, 2014.

\bibitem{SIMON1998151}
J.-M. Simon, D.K. Dysthe, A.H. Fuchs, and B.~Rousseau.
\newblock Thermal diffusion in alkane binary mixtures: A molecular dynamics
  approach.
\newblock {\em Fluid Phase Equilibria}, 150-151:151--159, 1998.

\bibitem{Perronace:2002aa}
A.~Perronace, G.~Ciccotti, F.~Leroy, A.H.~Fuchs, and B.~Rousseau,
\newblock Soret coefficient for liquid argon-krypton mixtures via equilibrium
  and nonequilibrium molecular dynamics: A comparison with experiments.
\newblock {\em Physical Review E}, 66(3), 2002.

\bibitem{Galli_ro_2003}
G.~Galli{\'{e}}ro, B.~Duguay, J.-P. Caltagirone, and F.~Montel.
\newblock On thermal diffusion in binary and ternary Lennard-Jones mixtures by
  non-equilibrium molecular dynamics.
\newblock {\em Philosophical Magazine}, 83(17-18):2097--2108, jan 2003.

\bibitem{YEGANEGI2005171}
Saeed Yeganegi and Masoud~Darvish Ganji.
\newblock Dependence of thermal diffusion factor of binary mixtures to the
  thermodynamic state by nemd simulation.
\newblock {\em Chemical Physics}, 318(3):171--179, 2005.

\bibitem{Evans_1991}
D.J. Evans and P.T. Cummings.
\newblock Non-equilibrium molecular dynamics algorithm for the calculation of
  thermal diffusion in simple fluid mixtures.
\newblock {\em Molecular Physics}, 72(4):893--898, 1991.


\bibitem{Heyes_1992}
D.~M. Heyes.
\newblock Molecular dynamics simulations of liquid binary mixtures: Partial
  properties of mixing and transport coefficients.
\newblock {\em The Journal of Chemical Physics}, 96(3):2217--2227, 1992.

\bibitem{Evteev_2014}
Alexander~V. Evteev, Elena~V. Levchenko, Irina~V. Belova, Rafal Kozubski,
  Zi-Kui Liu, and Graeme~E. Murch.
\newblock Thermotransport in binary system: case study on Ni$_50$Al$_50$ melt.
\newblock {\em Philosophical Magazine}, 94(31):3574--3602, 2014.

\bibitem{TUCKER201654}
William~C. Tucker and Patrick~K. Schelling.
\newblock Thermodiffusion in liquid binary alloys computed from
  molecular-dynamics simulation and the Green-Kubo formalism.
\newblock {\em Computational Materials Science}, 124:54--61, 2016.

\bibitem{Hafskjold_2017}
Bj{\o}rn Hafskjold.
\newblock Non-equilibrium molecular dynamics simulations of the transient
  ludwig-soret effect in a binary Lennard-Jones/spline mixture.
\newblock {\em The European Physical Journal E}, 40(1):4, 2017.

\bibitem{Bonella_2017} 
Sara Bonella, M. Ferrario, and G. Cicotti, 
\newblock Thermal diffusion in binary mixtures: Transient behavior and transport coefficients from equilibrium and nonequilibrium molecular dynamics.
\newblock{\em Langmuir}, 33(42): 11281, 2017.

\bibitem{Ferrario_2016}
M.~Ferrario, S.~Bonella, and G.~Ciccotti.
\newblock On the establishment of thermal diffusion in binary Lennard-Jones
  liquids.
\newblock {\em The European Physical Journal Special Topics},
  225(8-9):1629--1642, jul 2016.

\bibitem{Fernando_2020}
Kevin~M. Fernando and Patrick~K. Schelling.
\newblock Non-local linear-response functions for thermal transport computed
  with equilibrium molecular-dynamics simulation.
\newblock {\em Journal of Applied Physics}, 128(21):215105,  2020.

\bibitem{Bohm:2022aa}
Nathaniel Bohm and Patrick~K. Schelling
\newblock Analysis of ballistic transport and resonance in the
$ \alpha$-Fermi-Pasta-Ulam-Tsingou model.
\newblock {\em Physical Review E}, 106(2), 2022.

\bibitem{Parrinello_1981}
M.~Parrinello and A.~Rahman.
\newblock Polymorphic transitions in single crystals: A new molecular dynamics
  method.
\newblock {\em Journal of Applied Physics}, 52(12):7182--7190,  1981.

\bibitem{Bearman_1958}
Richard~J. Bearman and John~G. Kirkwood.
\newblock Statistical mechanics of transport processes. {XI}. Equations of
  transport in multicomponent systems.
\newblock {\em The Journal of Chemical Physics}, 28(1):136--145, 1958.

\bibitem{Hoang_2013}
Hai Hoang and Guillaume Galliero.
\newblock Local shear viscosity of strongly inhomogeneous dense fluids: from
  the hard-sphere to the lennard-jones fluids.
\newblock {\em Journal of Physics: Condensed Matter}, 25(48):485001, 2013.

\bibitem{GALLIERO2003171}
Guillaume Galli{\'e}ro, Bernard Duguay, Jean-Paul Caltagirone, and Fran{\c
  c}ois Montel.
\newblock Thermal diffusion sensitivity to the molecular parameters of a binary
  equimolar mixture, a non-equilibrium molecular dynamics approach.
\newblock {\em Fluid Phase Equilibria}, 208(1):171--188, 2003.

\bibitem{Bosse:1986vu}
J.~Bosse, G.~Jacucci, M.~Ronchetti, and W.~Schirmacher.
\newblock Fast sound in two-component liquids.
\newblock {\em Physical Review Letters}, 57(26):3277--3279, 1986.

\bibitem{Schram_1990}
R.~P.~C. Schram, A.~Bot, H.~M. Schaink, and G.~H. Wegdam.
\newblock Fast and slow sound in binary fluid mixtures.
\newblock {\em Journal of Physics: Condensed Matter}, 2(S):SA157--SA160, 
  1990.

\bibitem{Howard_1964}
R~E Howard and A~B Lidiard.
\newblock Matter transport in solids.
\newblock {\em Reports on Progress in Physics}, 27(1):161--240,  1964.

\bibitem{deGroot_1952}
S.~R. de~Groot.
\newblock {\em Thermodynamics of irreversible processes}.
\newblock Amersterdam: North Holland, 1952.

\bibitem{kohler_book}
W.~K{\"o}hler and S.~Wiegand, editors.
\newblock {\em Thermal Nonequilibrium Phenomena in Fluid Mixtures}, volume 584
  of {\em Lecture Notes in Physics}.
\newblock Springer Berlin Heidelberg, 2014.

\bibitem{Koehler_1995}
W.~K{\"o}hler and P.~Rossmanith.
\newblock Aspects of thermal diffusion forced rayleigh scattering: Heterodyne
  detection, active phase tracking, and experimental constraints.
\newblock {\em The Journal of Physical Chemistry}, 99(16):5838--5847, apr 1995.

\bibitem{Perronace_2002}
Andrea Perronace, Cindy Leppla, Fr{\'{e}}d{\'{e}}ric Leroy, Bernard Rousseau,
  and Simone Wiegand.
\newblock Soret and mass diffusion measurements and molecular dynamics
  simulations of n- pentane--n-decane mixtures.
\newblock {\em The Journal of Chemical Physics}, 116(9):3718--3729, mar 2002.

\bibitem{Debuschewitz:2001aa}
C.~Debuschewitz and W.~K{\"o}hler.
\newblock Molecular origin of thermal diffusion in benzene + cyclohexane
  mixtures.
\newblock {\em Physical Review Letters}, 87(5), 2001.

\bibitem{Duhr_2006}
Stefan Duhr and Dieter Braun.
\newblock Why molecules move along a temperature gradient.
\newblock {\em Proceedings of the National Academy of Sciences},
  103(52):19678--19682, dec 2006.

\end{thebibliography}

\end{document}